\documentclass{emulateapj}
\shorttitle{Polarimetry of NGC~891}
\shortauthors{Montgomery et al.}
\usepackage{graphicx,amsmath,epstopdf,natbib}
\begin{document}
\title{NEAR-INFRARED POLARIMETRY OF THE \mbox{EDGE-ON} GALAXY NGC~891}

\author{J. D. Montgomery and D. P. Clemens}
\affil{Boston University, Institute for Astrophysical Research
    725 Commonwealth Ave. Boston, MA 02215}
\email{montgojo@bu.edu}

\begin{abstract} 
The \mbox{edge-on} galaxy NGC~891 was probed using near-infrared (NIR) imaging
polarimetry in the $H$-band (1.6~$\mu$m) with the Mimir instrument on the 1.8~m
Perkins Telescope. Polarization was detected with signal-to-noise ratio greater
than three out to a surface brightness of 18.8~mag~arcsec$^{-2}$. The unweighted
average and dispersion in polarization percentage ($P$) across the full disk
were 0.7\% and 0.3\%, respectively, and the same quantities for polarization
position angle (P.A.) were 12\arcdeg\ and 19\arcdeg, respectively. At least one
polarization null point, where $P$ falls nearly to zero, was detected in the NE
disk but not the SW disk. Several other asymmetries in $P$ between the northern
and southern disk were found and may be related to spiral structure. Profiles of
$P$ and P.A. along the minor axis of NGC~891 suggest a transition from magnetic
(B) field tracing dichroic polarization near the disk mid-plane to scattering
dominated polarization off the disk mid-plane. A comparison between NIR P.A. and
radio (3.6~cm) synchrotron polarization P.A. values revealed similar
\mbox{B-field} orientations in the central-northeast region, which suggests that
the hot plasma and cold, star-forming interstellar medium may share a common
\mbox{B-field}. Disk-perpendicular polarizations previously seen at optical
wavelengths are likely caused by scattered light from the bright galaxy center
and are unlikely to be tracing poloidal \mbox{B-fields} in the outer disk.
\end{abstract}

\keywords{polarization---infrared: galaxies---galaxies: magnetic
fields---galaxies: individual (NGC~891)}

\section{Introduction}

Viewed externally and \mbox{edge-on}, how would the magnetic (B) field threading
the cold interstellar medium (ISM) of the Milky~Way appear? How is the ISM
\mbox{B-field} generated and sustained, and what is its relationship to the
radio synchrotron traced \mbox{B-field} in the hot plasma? NGC~891 is a bright,
nearby \citep[$d = 9.1$~Mpc,][]{2011ApJS..195...18R}, \mbox{edge-on} \citep[$i =
89.8\arcdeg$,][]{1998A&A...331..894X, 2005MNRAS.358..481K}, Milky~Way analog
ideally suited and situated for answering these questions.

Polarization studies and other methods have revealed that \mbox{B-fields}
permeate the ISM of the Milky Way and other galaxies \citep{1949ApJ...109..471H,
1949Sci...109..165H,1949Sci...109..166H, 1996Natur.379...47B,
1997A&A...318..700B, 2013AJ....145...74C}. Radio synchrotron emission is
partially linearly polarized, with its sky projected orientation being
perpendicular to the \mbox{B-field} in the hot ISM. Thus, synchrotron
polarization has been used extensively to study the \mbox{B-field} associated
with the hot plasma of external galaxies \citep{1996ARA&A..34..155B}. However,
long radio wavelengths suffer Faraday rotation and depolarization, and
accurately correcting for these is difficult \citep{1998MNRAS.299..189S,
1999MNRAS.303..207S}. Many \mbox{edge-on} galaxies studied in the radio show a
characteristic \mbox{X-shaped} polarization morphology, where the \mbox{B-field}
turns from disk-parallel to disk-perpendicular in the outer halo
\citep{1994A&A...284..777G, 2006A&A...448..133K}. Radio studies of NGC~891 have
revealed such an X-shape \citep{1991ApJ...382..100S, 2009RMxAC..36...25K}.

Background optical and near-infrared (NIR) starlight probes \mbox{B-fields} in
the dusty ISM. Radiative torques align aspherical dust grains with the long axis
mostly perpendicular to the \mbox{B-field} \citep{2007MNRAS.378..910L}. This net
alignment causes dichroic extinction, which linearly polarizes background
starlight to exhibit an orientation parallel to the plane-of-sky \mbox{B-field}
projection. Dichroic polarization percentage increases with optical depth,
though not always linearly \citep{1989ApJ...346..728J, 1992ApJ...389..602J}.

In addition to dichroic polarization, another source of polarization is light
singly (and less so, multiply) scattered by dust grains. This light is
polarized perpendicular to the plane of scattering and rarely relates to the
\mbox{B-field} orientation. Therefore, it is important to distinguish between
dichroism and scattering in NIR and optical studies of \mbox{B-field}s to avoid
incorrect inferences regarding the \mbox{B-field} morphology and other
properties.

Although dust grain scattering cross-sections are inversely proportional to
wavelength \citep{1983asls.book.....B}, the relationship between scattering
polarization percentage and wavelength remains unclear.
\citet{1992ApJ...400..238S} studied the reflection nebulae NGC~7023 and NGC~2023
and found no significant difference between $V$- and $H$-band polarization
percentage. Thus, NIR wavelengths can probe deeper into the disk of NGC~891 than
optical wavelengths, but scattering may still be an important component of the
light and a possible contaminant in NIR \mbox{B-field} studies.

Modern, high sensitivity, wide-field NIR polarimetry instruments are now able to
probe \mbox{B-fields} across entire galaxies \citep{2007PASP..119.1385C,
2012ApJ...761L..28P}. The $H$-band polarization of the central 60~arcsec of
NGC~891 was measured by \citet{1997AJ....114.1393J}, who found weak
polarizations with generally disk-parallel orientation. NIR polarimetry of
\emph{entire}, Milky~Way analog disks will provide context for detailed studies
of the Milky~Way \mbox{B-field} \citep{1970MmRAS..74..139M,
2012ApJS..200...19C}.

This current study sought to obtain $H$-band observations that are sufficiently
sensitive to measure the polarization at arcsecond resolution out to nearly the
full extent of the disk of NGC~891. These are the deepest NIR polarization
observations of an entire \mbox{edge-on} galaxy to date.
The observations revealed about 1\% polarization across the entire disk of
NGC~891. This $P$ value is weaker than the \citet[][hereafter
JKD]{1992ApJ...389..602J} model prediction for NGC~891. Polarization ``null
points,'' where $P$ falls nearly to zero, perhaps related to the null points
predicted by \citet{1997AJ....114.1405W} in their model of polarized galaxy
light, were detected. However, the observed null points are not located
symmetrically about the galaxy center, as was predicted. Therefore, the null
points are not due to radiative transfer effects alone but may relate to galaxy
structure such as spiral arms.

There is strong polarization for positions well off the disk mid-plane, across
the entire extent of NGC~891. This difference between mid-plane and off-plane
polarizations might be due to disk starlight scattered by the extraplanar dust
known to be present \citep{2000AJ....119..644H}.

The new NIR polarimetric observations and the data reduction procedures are
described in Section~\ref{sec:observations}. In Section~\ref{sec:analysis},
general polarization trends along the major and minor axes of NGC~891 are
analyzed, polarization maps are presented, and radio and NIR P.A.s are compared.
Section~\ref{sec:discussion} discusses how galaxy structure may influence
polarization, compares these observations to polarization radiative transfer
models, and compares the NIR and optical polarization properties.

\section{Observations and Data Reduction} \label{sec:observations}
NGC~891 was observed for $H$-band (1.6~$\mu$m) linear polarization over 15
nights between 2011 September 19 and October 20, using the Mimir instrument
\citep{2007PASP..119.1385C} on the Perkins 1.8~m telescope, located outside
Flagstaff, AZ. Mimir used an Aladdin III, $1024 \times 1024$~pixel, InSb array
detector, and had an equatorially aligned $10 \times 10$~arcmin field of view
(FOV), sampled at 0.58~arcsec~per~pixel.

NIR linear polarization was measured using a stepping, cold, compound half-wave
plate (HWP) and a cold, fixed wire grid. Images were obtained in a six-point
hexagonal dither pattern with 15~arcsec separations. At each pointing, 10~sec
exposures were obtained for each of 16 unique HWP position angles. Thus, each
exposure set consisted of 96 separate images. A total of 47 such sets were
collected, for a net exposure time of 12.5 hours covering the entire disk of
NGC~891.

The data were processed using the Mimir Software Package Basic Data Processing
(\mbox{MSP-BDP} v3.1) and Photo POLarimetry (PPOL v8.0) reduction tools. BDP
performed linearity correction, flat-fielding (specific to each HWP angle), dark
current correction, and returned science-quality images. These images were
processed by PPOL, which performed astrometry, corrected for time-varying sky
transmission, combined images taken through the same HWP angles, rejecting bad
or missing pixels, corrected for instrumental polarization across the FOV, and,
for stellar fields, also performed PSF-assisted aperture photometry
\citep{2012ApJS..200...20C}.

\subsection{Background Removal}
PPOL normally removes background sky flux by forming super-sky images, but this
method removes some of the object flux for extended objects larger than the
15~arcsec dither steps. Thus, for extended fields, sky correction was done by
modeling the background shape across the detector.

A $25 \times 25$ grid of uniformly spaced sample zones ($\sim$40 pixels)
far from the disk of NGC~891 was established for the first, provisional coadded
image. The mean background sky brightness within each zone was found for each of
the 96 images in an exposure set. Second-order, 2-D polynomials were fit to each
image's grid of average sky brightness values and subtracted from each image
prior to stacking, to obtain the background-free surface brightness. This
procedure was repeated for all 47 exposure sets.

Low-level, systematic trends remaining along the northern and southern edges of
the field of view were also removed. The average sky-brightness in each HWP
position angle image was measured in regions far from the galaxy and detector
edges. For each row in an image, the difference between the median filtered mean
brightness of that row and the 2-D corrected average sky brightness was
subtracted from the row.

These background-corrected HWP images were combined into mean images for each of
the 16 HWP angles. These were further combined to produce images of Stokes $I$,
$U$, and $Q$ and their respective uncertainties, where $U$ and $Q$ were
normalized by the Stokes intensity. Finally, the 47 images of each Stokes value
were astrometrically registered and averaged to achieve the highest sensitivity.

\subsection{Smoothing and Correction}
The averaged, normalized Stokes $U$ and $Q$ and uncertainty images were used to
compute $P$, P.A., and $\sigma_{\rm P.A.}$ images by the formulae
\begin{equation} \label{eq:pol}
	\begin{aligned}
		P                 &= \sqrt{U^2 + Q^2} \text{,} \\
		{\rm P.A.}        &= 0.5 \arctan\left(U/Q\right) \text{, and} \\
		\sigma_{\rm P.A.} &= 0.5 \arctan \left(\sigma_{P}/P\right) \text{,}
	\end{aligned}
\end{equation}
where $\sigma_{P}$ is defined by the error propagation of $\sigma_{U}$ and
$\sigma_{Q}$, and P.A. is measured east of north. However, these polarization
images exhibited high spatial frequency noise over most of the galaxy.
Therefore, the combined $U$ and $Q$ images were weighted by their respective
inverse variance images, and the weighted images were convolved with a gaussian
kernel. This achieved lower noise, though at the expense of some angular
resolution. To prevent a few low variance Stokes values (in high-flux pixels)
from dominating weighted means, variances used for weighting were trapped at a
lower limit of two standard deviations below the average Stokes variance.

Uncertainties in $U$ and $Q$ measurements positively bias $P$ values computed as
per Equation~\ref{eq:pol}. Therefore, unless otherwise noted, all reported $P$
values were Ricean corrected \citep{1974ApJ...194..249W} using the following
formula to remove the bias:
\begin{equation} P_{cor} =
	\begin{cases}
		\sqrt{P - \sigma_{P}^{2}}, &\text{if } P \ge \sigma_{P} \\
		0,                         &\text{if } P   < \sigma_{P}.
	\end{cases}
\end{equation}

\section{Analysis} \label{sec:analysis}
Polarization percentage and position angle images were examined for significant
spatial structure. These characteristics were explored by creating $P$, P.A.,
polarized intensity ($PI$), and intensity profiles along the major and minor
axes of NGC~891. Finally, NIR and radio P.A. maps were compared to test the
degree to which they probe the same \mbox{B-field}.

\subsection{Intensity and Polarization Maps}
\begin{figure}
{\centering
\includegraphics[angle=0, scale=1.8]
{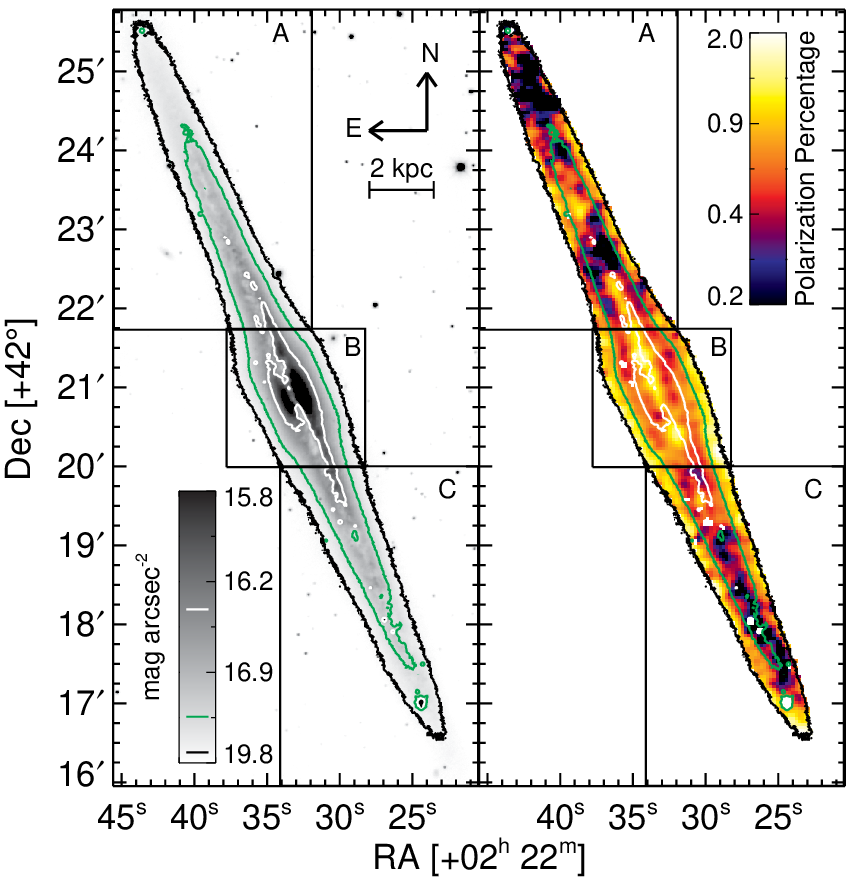}
\caption{
(left:) Gray-scale, linear representation of the $H$-band total surface
brightness distribution for NGC~891, overlaid with contours at 18.8~(black),
17.6~(green), and 16.4~mag~arcsec$^{-2}$~(white). (right:) Log-scaled and masked
map of polarization percentage. Overlaid intensity contours are the same as in
the left panel; all pixels outside the black contour are masked. The boundaries
of the regions A, B, and C, described in the text, are drawn and labeled in
black.
\label{fig:I_Pol}}}
\end{figure}

The unsmoothed intensity image, overlaid with surface brightness contours at
18.8 (black), 17.6 (green), and 16.4~mag~arcsec$^{-2}$ (white), is shown in the
left panel of Figure~\ref{fig:I_Pol}. Labeled boxes delineate the three distinct
galaxy regions analyzed later in this paper: (A) the northeast (NE) disk, (B)
the central disk and bulge, and (C) the southwest (SW) disk. The image exhibits
stellar point spread functions (PSF) with full width at half maxima (FWHM) of
1.75~arcsec. Surface brightness was calibrated using 2MASS stellar photometry,
and a mean background sky brightness of 21.4~mag~arcsec$^{-2}$ was found.

Although mostly symmetric, each surface brightness contour in
Figure~\ref{fig:I_Pol} reveals an asymmetry in the light distribution from
NGC~891. The black contour extends 30~arcsec (1.5~kpc) further to the NE than to
the SW. The extreme SW end of the green contour reveals a deficit of light from
the western side of the major axis, while the white contour in the central bulge
reveals more light from the western side of the major axis than from the eastern
side. This last asymmetry is caused by the slightly less than 90\arcdeg\
inclination and indicates the inclination direction.

The right panel of Figure~\ref{fig:I_Pol} shows the polarization map overlaid
with the same surface brightness contours. This map was generated by smoothing
the $U$ and $Q$ images with a 4.5~arcsec FWHM kernel and resampling with
2.4~arcsec pixel spacing prior to forming the $P$ image. The dust lane in the
central region shows strong $P$, which decreases in strength just above and
below the disk mid-plane and returns to $P > 1\%$ further from the mid-plane.
Polarization percent outside the central region is generally weak, less than
$1\%$, except in a few locations.

The NE disk shows two regions of very low polarization, here termed inner and
outer ``null points,'' at 42\arcdeg 22\arcmin 45\arcsec\ and 42\arcdeg 24\arcmin
45\arcsec\ decl. Polarization percentage in the SW disk is weak along the dust
lane but strong ($P > 1\%$) far from the disk mid-plane. The strip of low
polarization along the SW dust lane begins about 120~arcsec (5.3~kpc) from the
galaxy center, which is about the same offset angle as for the NE inner null
point.

\subsection{Polarization Profiles}
Are the null points observed in Figure~\ref{fig:I_Pol} significant? What are the
trends in polarization along the major and minor axes? What unique information
is contained in each of the $P$, P.A., $PI$, and intensity quantities? These
questions were addressed using profiles of these quantities along the major and
minor axes.

The major axis of NGC~891 was identified as the ridge of bright 24~$\mu$m
emission observed by the MIPS instrument on the \textit{Spitzer Space Telescope}
\citep{2012MNRAS.423..197B}. When seeking the minor axis, the 24~$\mu$m image
did not show a bright galaxy center, so it was instead astrometrically
registered with the Mimir $H$-band image using SAO DS9. The minor axis locus was
then defined to be the line perpendicular to the major axis that passed through
the brightest location of the $H$-band image. In the following profile analyses,
the positive major and minor axis directions were defined to be toward the
southwest and northwest, respectively. The major and minor axis positions of all
pixels within the $H$-band 18.8~mag~arcsec$^{-2}$ isophot in the unsmoothed
images were computed, and the profiles were generated as described below.

The major axis profiles aimed to probe changes in \mbox{B-field} properties
rather than effects from scattering. Based upon the minor axis profile discussed
below, the mid-plane region appears to be dominated by dichroism, while light
far from the mid-plane region is dominated by scattering. Therefore, the major
axis pixels between minor axis offsets of $-9$~arcsec and $+6$~arcsec, where
scattering effects are weakest, were selected for inclusion in the major axis
profile. The pixels within this region were grouped into bins 5~arcsec wide
along the major axis, and inverse variance weighted $U$ and $Q$ averages and an
unweighted intensity average were computed for each bin, along with
uncertainties in those means. Mean $U$ and $Q$ values were substituted into
Equation~\ref{eq:pol} to compute a $P$ profile, and the $PI$ profile was
computed as the product of the $P$ and intensity profiles. All profiles were
folded about the minor axis to compare behavior in the north and the south.

For the minor axis profiles, three non-contiguous regions, extending 100~arcsec
along the major axis, were defined to distinguish polarization behavior in
different parts of the disk. These regions were centered on the galaxy nucleus
and at 150~arcsec offsets to the NE and SW, which roughly correspond to the
centers of the A, B, and C regions marked in Figure~\ref{fig:I_Pol}.

These profiles aimed to probe variations in the polarization along the minor axis
direction. Therefore, pixels inside the northern inner null point, defined to be
the region where $P \leq 0.2\%$, were excluded from the weighted means. These
pixels constituted as much as $15\%$ of the data in the bins between 2--8 arcsec
in the NE disk minor axis profile, so the null point significantly influenced
the weighted means at those locations, despite the high variances inside the
null point. The remaining pixels within the regions described above were grouped
into bins 1.5~arcsec wide along the minor axis. Inverse variance weighted $U$
and $Q$ means and uncertainties in the mean were computed for each bin in each
of the three regions. Profiles of $P$ and P.A. were computed by substituting
mean $U$ and $Q$ values into Equation~\ref{eq:pol}.

\subsubsection{Major Axis Profiles}
\begin{figure}
{\centering
\includegraphics[angle=0, scale=1.2]
{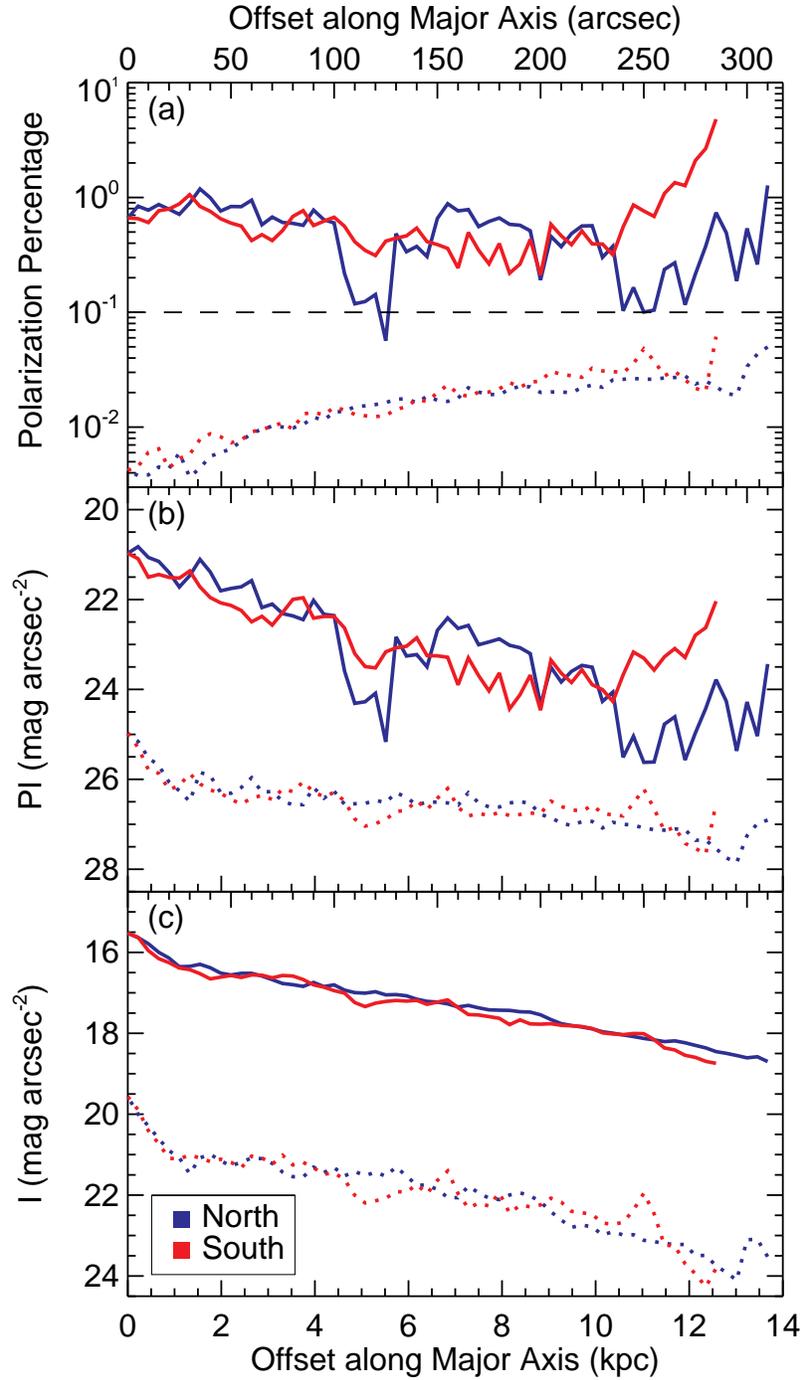}
\caption{
(a, top:) NGC~891 major axis profiles of averaged polarization percentage (solid
lines) and associated uncertainties (dotted lines), versus angular and physical
offsets along the northern (blue lines) and southern (red lines) disk regions.
The horizontal dashed line at 0.1\% indicates the Mimir calibration (external
uncertainty) limit. The profile is relatively flat except for dips (``nulls'')
in the north at 5.5~kpc and 11~kpc offsets. In contrast, the south shows no
strong dips in the $P$ profile. (b, middle:) Corresponding major axis profiles
for the polarized intensity and uncertainties. (c, bottom:) Major axis profiles
for the $H$-band intensity. The bright core (offsets less than 30~arcsec or
1.5~kpc) shows a steeper decrease in brightness than the rest of the disk.
\label{fig:majorPro}}}
\end{figure}

The major axis profiles of $P$, $PI$, and intensity are shown in
Figure~\ref{fig:majorPro}. These $P$ values were \emph{not} Ricean corrected.
Solid lines represent the averaged quantities while dotted lines represent their
internal, statistical uncertainties. A dashed line at $P = 0.1\%$ in panel (a)
marks the systematic, external calibration limit of the Mimir instrument
\citep{2012ApJS..200...20C}.

The $P$ profile, shown in Figure~\ref{fig:majorPro}.a, is mostly flat as a
function of offset, except for the null points in the north, which are offset
from the galaxy center by 125~arcsec (5.5~kpc) and 250~arcsec (11~kpc), and for
which the $P$ values drop from $\sim$0.6\% to $\sim$0.1\%. Beyond
280~arcsec (12.4~kpc), $P$ returns to $\sim$0.5$\pm$0.1\%. Weak
contamination from systematic edge-effects may exist, so the inner null point
represents a more confident detection than the outer null point. There is an
exponential rise in $P$ in the extreme south, with a slope of
$\sim$0.4\%~kpc$^{-1}$. If this feature is real, then the extreme SE disk
is the \emph{most} polarized part of the galaxy along the major axis.

The intensity profile (Figure~\ref{fig:majorPro}.c) shows a bright nucleus,
beyond which there is an exponential decrease in brightness with a slope of
about 0.13~mag~arcsec$^{-2}~$kpc$^{-1}$. The $PI$ profile
(Figure~\ref{fig:majorPro}.b) also decreases with offset from the galaxy center
because of the corresponding decrease in intensity. All sub-kpc scale structure
in the $PI$ profile originates with the $P$ profile. Thus, these major axis
profiles show that, in the NIR $H$-band, $PI$ carries no unique information.
Furthermore, unlike in the radio, the ratio of $PI$ and $I$ in the NIR is
unrelated to magnetic field strength.

\subsubsection{Minor Axis Profiles}
\begin{figure}
{\centering
\includegraphics[angle=0, scale=1.65]
{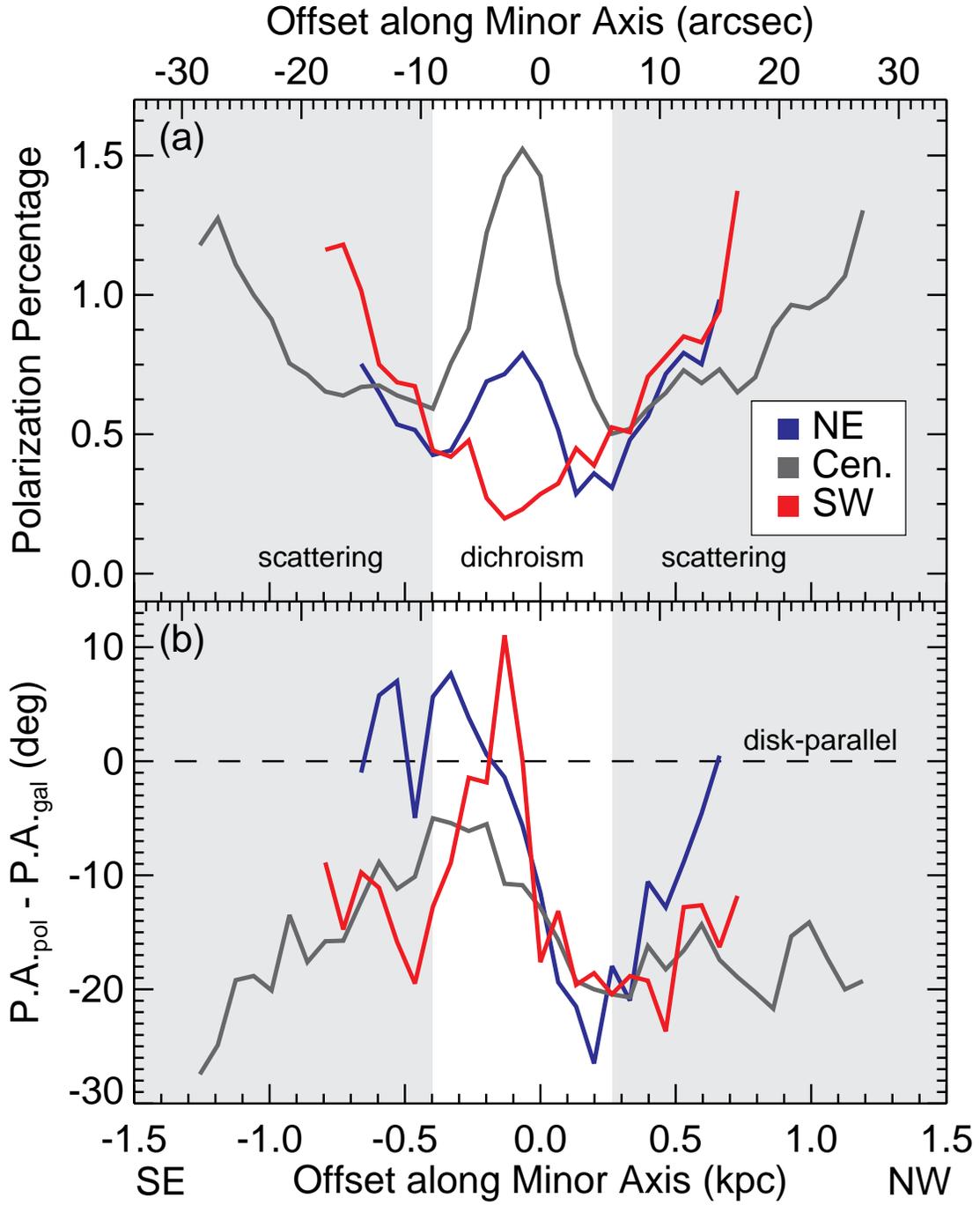}
\caption{
Profiles of $P$ and disk-relative P.A. along the minor axis of NGC~891. (a,
top:) The NE disk and central bulge peak in $P$ near $-0.1$~kpc, while the SW
disk has a $P$ minimum there. (b, bottom:) There is a swing from disk-parallel
P.A., at $-0.1$~kpc, to $-20$\arcdeg\ offset in P.A. at $+0.2$~kpc.
Uncertainties in these profiles are discussed in the text.
\label{fig:minorPro}}}
\end{figure}

The minor axis $P$ and P.A. profiles for the A, B, and C regions are shown in
Figure~\ref{fig:minorPro}. The galaxy major axis P.A. was subtracted from the
polarization P.A. values such that a value of zero represents a disk-parallel
polarization orientation, while positive difference P.A.s still represent
rotations east of the major axis.

The dispersions of normalized $U$ and $Q$ values in the minor axis bins were
between $\sim$0.005--0.06\%, i.e., very small. These dispersions and the roughly
$\sim$450 independent samples in the minor axis bins lead to internal
uncertainties in the mean $P$ and P.A. between 0.2--40$\times 10^{-3}$\% and
0.008--0.8\arcdeg, respectively (smaller than the vertical minor tick mark steps
in Figure~\ref{fig:minorPro}), though the calibration limit of $P = 0.1\%$ for
the Mimir instrument affects external camparisons.

The most prominent features in the Figure~\ref{fig:minorPro} profiles are
coincident with the dust lane, which lies immediately below the major axis, due
to the slightly less than \mbox{edge-on} inclination of NGC~891. In
Figure~\ref{fig:minorPro}.a, the central and NE regions show local maxima at
$-1.5$~arcsec, while the SW profile shows a local minimum. These features have
full-widths of about 10~arcsec (400~pc), which is more than six times greater
than the bin width. Far from the disk mid-plane, all three regions show strong
increases in $P$ with minor axis offset. Those increases show half-width scale
heights of 25~arcsec (1.1~kpc) in the NE and SW disk and are wider, 40~arcsec
(1.8~kpc), for the central region.

\citet{1997AJ....114.1393J} found that the typical polarization P.A. in the dust
lane of the central region was about 20\arcdeg\ to the west of the major axis
P.A. These new data confirm a similar P.A. offset across most of the central
region. Interestingly, all three regions show a shift from roughly disk-parallel
P.A. below (SE of) the dust lane to about $-20$\arcdeg\ P.A. above (NW of) the
dust lane. The P.A. values in the NE return to disk-parallel orientation further
from the dust lane while the SW and central region P.A. values at these same
locations are typically between $-15$\arcdeg\ and $-25$\arcdeg. This is the
first time a significant difference between the major axis P.A. and the NIR
polarization P.A.s has been measured in the NGC~891 outer disk, far from the
bulge.

\subsection{Polarizations toward the Disk and Bulge}
Where dichroism is the dominant polarizing mechanism, the polarization P.A.
reveals the plane-of-sky projection of the \mbox{B-field} threading the cold
dust in the disk \citep{1997ApJ...477L..25W}. Polarization vector maps were
plotted in order to examine the geometry of the \mbox{B-field}. In general,
$\alpha\Omega$~dynamo theory \citep[see review by][]{1988Natur.336..341R}
predicts the \mbox{B-field} in the disk should be predominantly toroidal,
producing disk-parallel polarizations. The data presented here for the disk
mid-plane of NGC~891 generally agree with these predictions. This result stands
in contrast to the optical polarizations for NGC~891, which show a nearly
disk-perpendicular orientation across most of the disk
\citep{1996MNRAS.278..519S}.

\begin{figure}
{\centering
\includegraphics[angle=0, scale=1.5]
{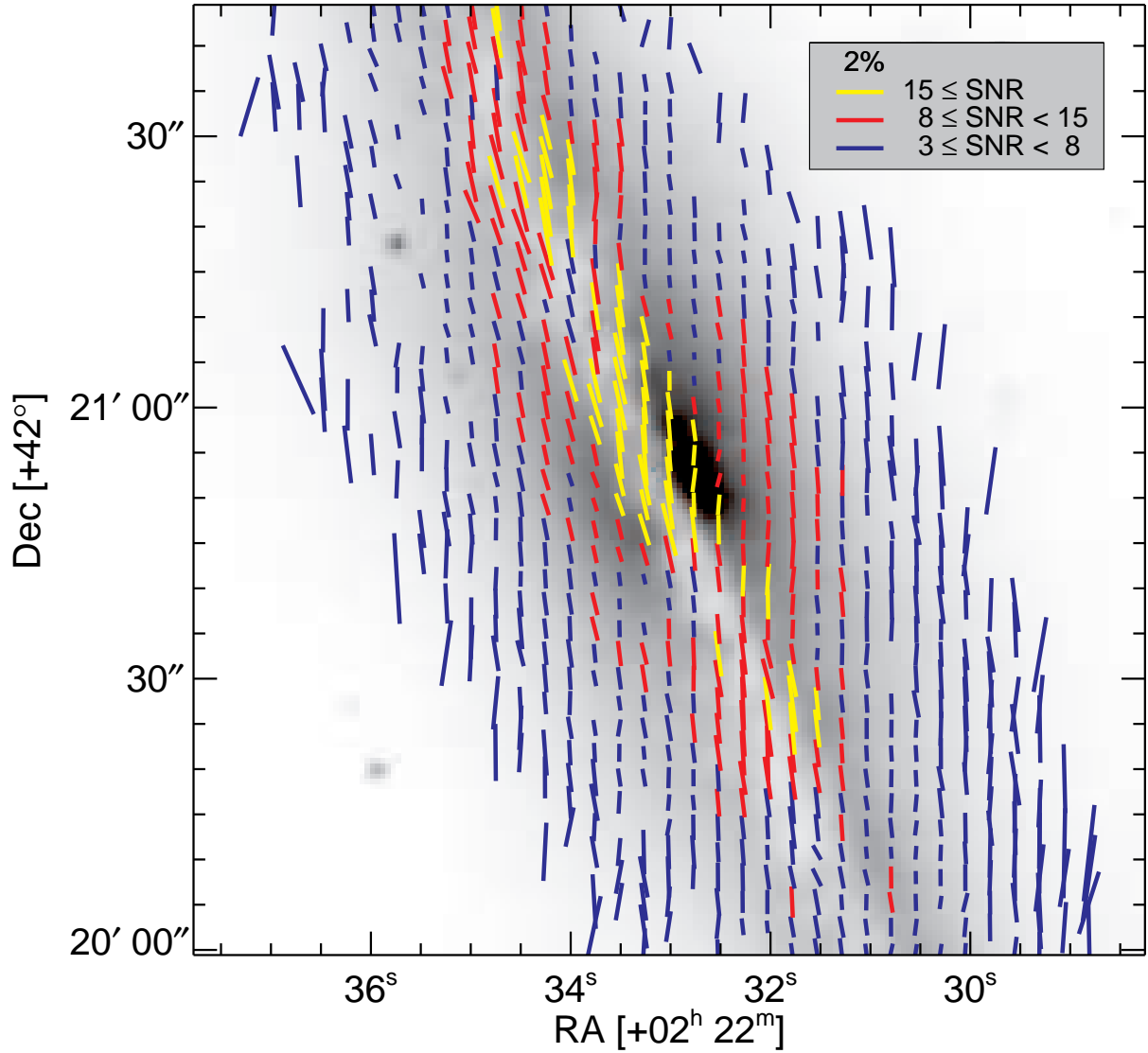}
\caption{
Gray-scale, linear, $H$-band intensity image for the central bulge, ``B'' region
in Figure~\ref{fig:I_Pol}, overlaid with smoothed, resampled polarization
vectors. Vector lengths and colors correspond to polarization percentage and
SNR, as indicated in the legend. \label{fig:regB}}}
\end{figure}

\begin{figure}
{\centering
\includegraphics[angle=0, scale=1.5]
{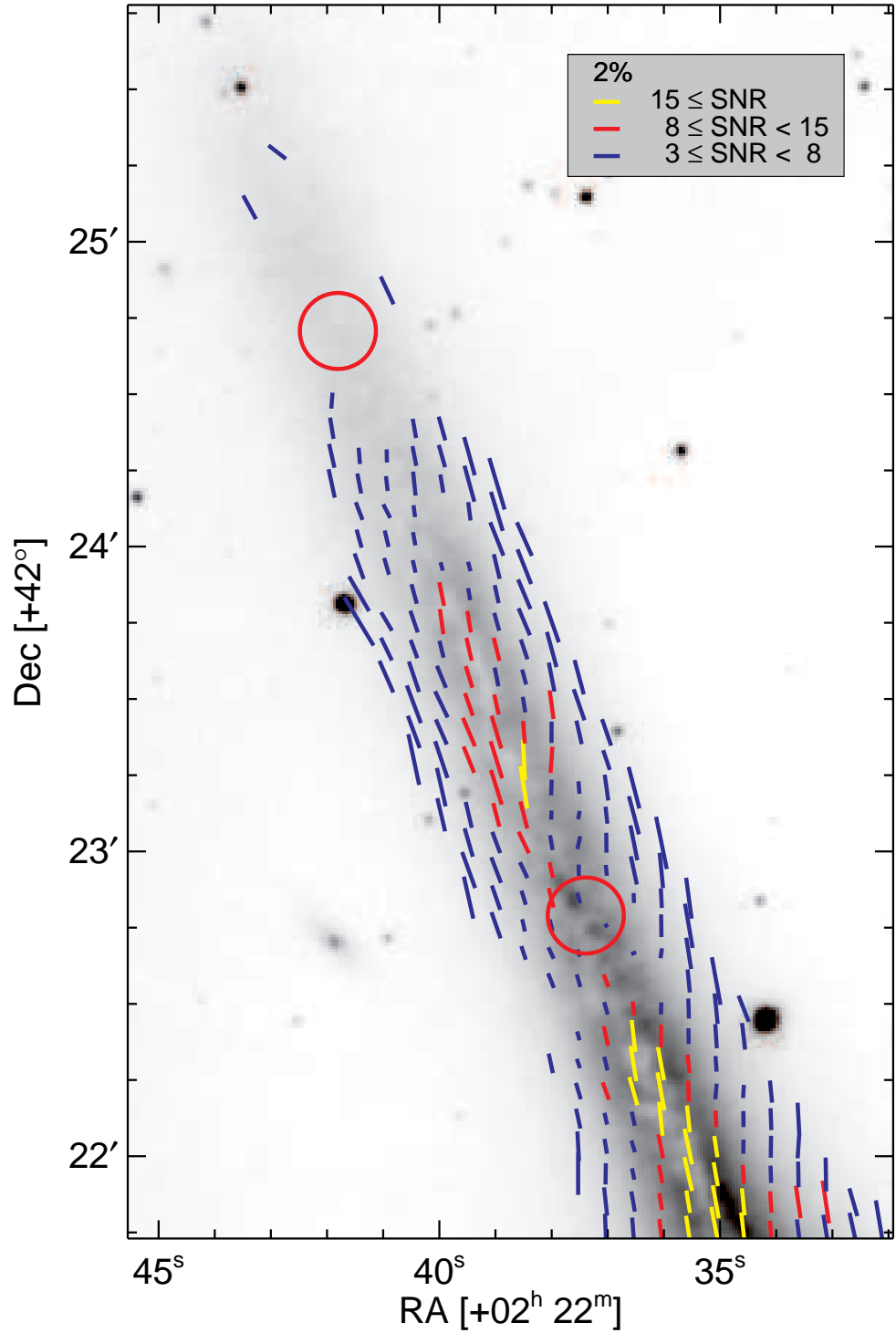}
\caption{
Gray-scale, linear, $H$-band intensity image for the northeast disk, ``A''
region of Figure~\ref{fig:I_Pol}, overlaid with smoothed, resampled polarization
vectors. The red circles along the disk mid-plane indicate the locations of the
polarization null points described in the text. Vector lengths and colors
correspond to polarization percentage and SNR, as indicated in the legend.
\label{fig:regA}}}
\end{figure}

\begin{figure}
{\centering
\includegraphics[angle=0, scale=1.5]
{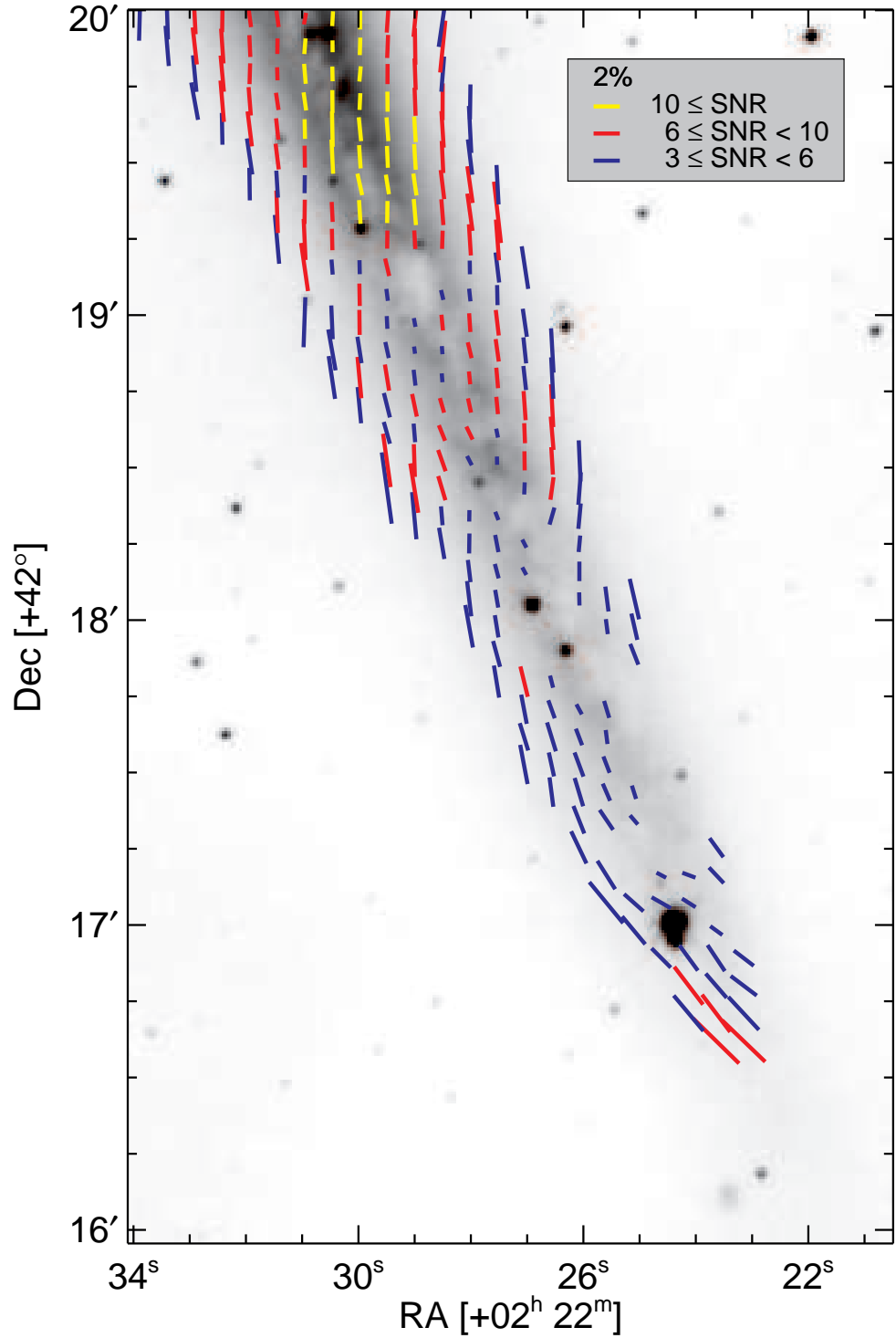}
\caption{
Gray-scale, linear, $H$-band intensity image for the southwest disk, ``C''
region of Figure~\ref{fig:I_Pol}, overlaid with smoothed, resampled polarization
vectors. Vector lengths and colors correspond to polarization percentage and
SNR, as indicated in the legend. The polarization percentage is lowest in this
region, so different SNR ranges were used in order to show the range of
polarization SNR across this region.
\label{fig:regC}}}
\end{figure}

For analysis, NGC~891 was divided into the three regions, labeled A, B, and C in
Figure~\ref{fig:I_Pol}. The polarization signal was strongest in the central
region B, so 1.7~arcsec FWHM gaussian smoothing could be applied there and still
retain high SNR, while a 4.5~arcsec FWHM smoothing was needed for regions A and
C. These kernels were applied to form $U$ and $Q$ images which were used to
produce $P$ and P.A. images, in which values with ${\rm SNR} < 3$ were masked.
The smoothed $U$ and $Q$ images were resampled with 2.7~arcsec pixel spacing to
produce $P$ and P.A. for region B, and 5.4~arcsec pixel spacing for regions A
and C, so that each displayed vector was effectively independent. Polarization
vectors corresponding to the resampled pixels were plotted over the $H$-band
intensity image and are shown in Figure~\ref{fig:regB}~(region B; central),
Figure~\ref{fig:regA}~(region A; NE disk), and Figure~\ref{fig:regC}~(region C;
SW disk). Vector lengths and orientations correspond to $P$ and P.A.,
respectively, while colors correspond to SNR, as noted in the legend of each
figure. Lower SNR ranges were used in Figure~\ref{fig:regC} because $P$ is
weakest in region C.

\citet{1997AJ....114.1393J} argued that his $H$-band polarizations above and
below the central disk turned disk-perpendicular to match the poloidal halo
\mbox{B-fields} observed in $\lambda$6.2~cm radio synchrotron by
\citet{1991ApJ...382..100S}. However, the vectors in
Figure~\ref{fig:regB}--\ref{fig:regC} are generally disk-parallel and have a
maximum P.A. uncertainty of only about 10\arcdeg. Given these low uncertainties,
the $H$-band polarizations do not indicate the presence of poloidal fields in
any part of the disk or off the disk mid-plane of NGC~891.

Polarizations in the NE disk (Figure~\ref{fig:regA}) are generally more
disk-parallel than in the other regions of NGC~891. Few measurements in the far
NE disk have ${\rm SNR} > 3$ because $P$ is weak there. However, the
polarization in this extreme end of the disk remains disk-parallel.

\citet{1990IAUS..140..245S} observed NGC~4565 in the optical and found a
transition from disk-parallel to disk-perpendicular polarization located about
75~arcsec (4~kpc) from the nucleus along the major axis.
\citet{1997AJ....114.1405W} reproduced this transition using the Monte-Carlo
radiative transfer model developed by \citet{1997ApJ...477L..25W}. They
predicted that the transition would be associated with a polarization null point
caused by the cancellation of orthogonal polarizations from bulge light
scattered off dust in the disk and dichroic polarization of background disk
starlight. They also predicted that the null point transition would appear
closer to the nucleus in NIR than in optical.

Figure~\ref{fig:majorPro} shows the first detection of any polarization null
point in NGC~891 (marked by red circles in the Figure~\ref{fig:regA} NE disk
image). However, there is no corresponding transition from disk-parallel to
disk-perpendicular polarizations across the inner null point in
Figure~\ref{fig:regA}, as would have been expected if the polarization dip were
caused by competition between scattering and dichroism. Therefore, these
represent a new class of polarization null points, ones not described by the
\citet{1997ApJ...477L..25W} model.

Much of the SW disk (Figure~\ref{fig:regC}) continues the 20\arcdeg\ westward
offset between major axis P.A. and polarization P.A. As seen in
Figures~\ref{fig:I_Pol} and \ref{fig:minorPro}.a, $P$ is weak through the SW
dust lane, which is indicated by the absent or short vectors along the dust lane
in Figure~\ref{fig:regC}. In the far SW disk, $P$ is strong and the polarization
P.A.s are offset from the major axis P.A. in the \emph{opposite} direction of
that seen elsewhere in the galaxy. This indicates that the \mbox{B-field} does
\emph{not} turn poloidal in the outer disk as was concluded by
\citet{1996MNRAS.278..519S} and \citet{1997AJ....114.1393J}.

\subsection{Comparison to Radio Data}
To date, the relationship between the \mbox{B-field} in the hot plasma and the
\mbox{B-field} in the cold ISM of external galaxies has not been observationally
constrained. However, comparing the polarization P.A. from synchrotron emission
and NIR dichroic polarizations for NGC~891 tests whether there is agreement
between these two \mbox{B-fields}.

Magnetic field P.A. values from polarized synchrotron emission at $\lambda3.6$~cm from
NGC~891 were extracted from Figure~1 of \citet{2009RMxAC..36...25K}. At this
wavelength, Faraday rotation is relatively weak and easily corrected, and the
half-power beamwidth (HPBW) of the Effelsberg 100~m telescope, used to collect
the $\lambda3.6$~cm data, was 84~arcsec. The NIR $U$ and $Q$ images were
smoothed with a 4.5~arcsec kernel and resampled with 9.6~arcsec pixel spacing to
ensure independence between NIR polarization values prior to comparison with the
radio values.

\subsubsection{P.A. Distributions}
\begin{figure}
{\centering
\includegraphics[angle=0, scale=1.5]
{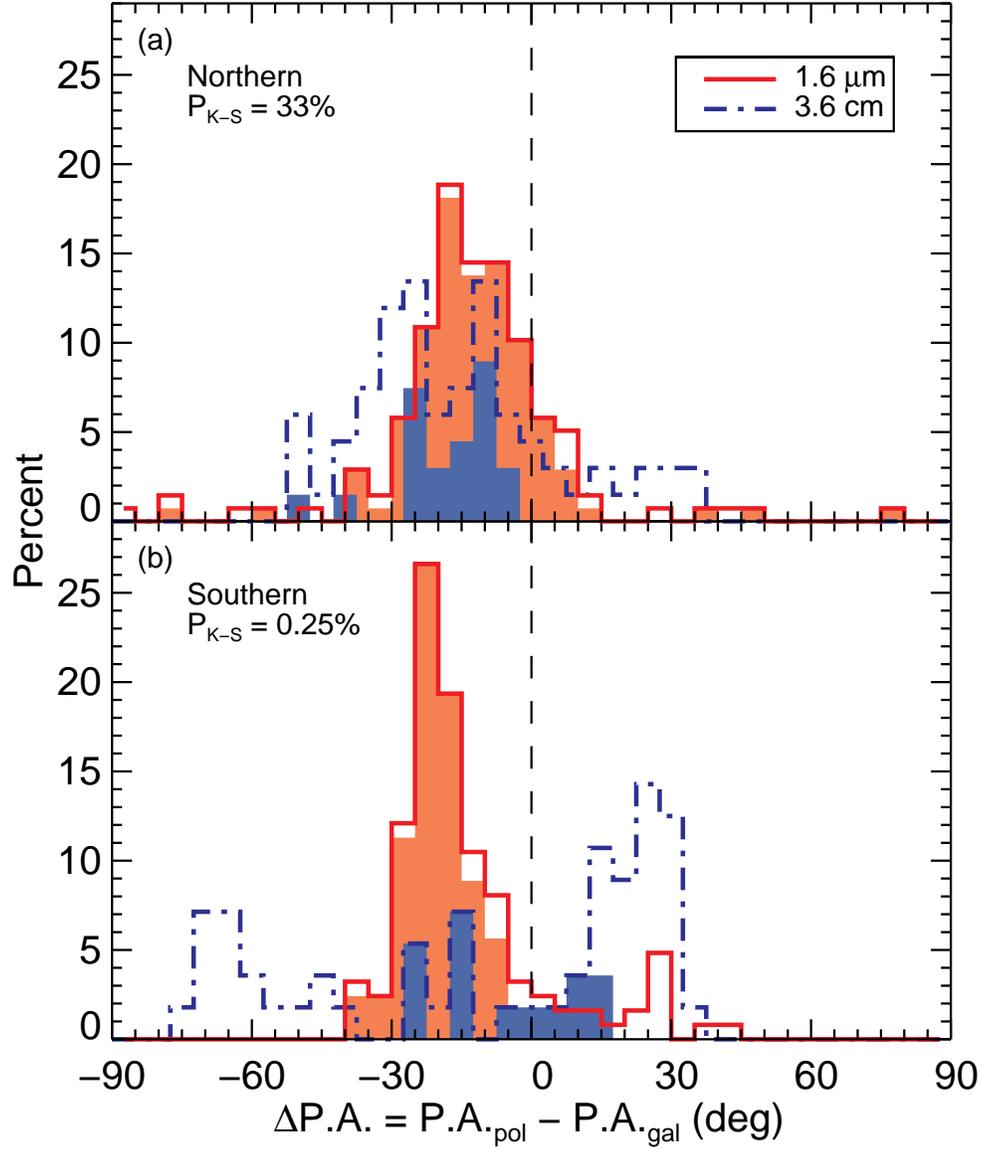}
\caption{
Histograms of NIR (solid, red lines) and radio (dot-dashed, blue lines)
polarization P.A. values relative to the P.A. of the major axis for the northern
(a, top) and southern (b, bottom) halves of NGC~891. The red and blue shaded
(filled) portions represent the measurements obtained within the solid angle
common to the radio map and the 18.8~mag~arcsec$^{-2}$ $H$-band isophot. The
major axis P.A. is marked by the dashed line at $\Delta{\rm P.A.} = 0$. The
Kolmogorov-Smirnov statistic reported in the top-left of each panel is the
probability that the two shaded distributions are drawn from the same parent
distribution.
\label{fig:histograms}}}
\end{figure}

The disk was divided about its minor axis into northern and southern halves, and
the galaxy major axis P.A. was subtracted from the NIR and radio polarization
P.A. values to form $\Delta{\rm P.A.}$ The resulting $\Delta{\rm P.A.}$ values
were grouped into 5\arcdeg\ wide bins, which is the typical uncertainty at
$\lambda$3.6~cm and $H$-band, so each bin should be largely independent. The
resulting $\Delta{\rm P.A.}$ distributions are shown in
Figure~\ref{fig:histograms}, where NIR distributions are indicated with solid,
red lines and the radio distributions are marked by dot-dashed, blue lines.

In the northern disk (Figure~\ref{fig:histograms}.a), the median and dispersion
of the NIR $\Delta{\rm P.A.}$ values are about $-15$ and $15$\arcdeg,
respectively. The corresponding median and dispersion of the radio $\Delta{\rm
P.A.}$ values are about $-20$ and $20$\arcdeg. In the southern disk
(Figure~\ref{fig:histograms}.b), the NIR $\Delta{\rm P.A.}$ distribution has
median and dispersion values of about $-20$ and  $10$\arcdeg, while the radio
distribution is trimodal, with groups of $\Delta{\rm P.A.}$ values at
$-65$\arcdeg, $-20$\arcdeg, and $+25$\arcdeg and dispersions of about 15\arcdeg,
7.5\arcdeg, and 10\arcdeg, respectively.

Measurements obtained within the solid angle common to the radio map and the
18.8~mag~arcsec$^{-2}$ $H$-band isophot were selected for detailed comparison
via a Kolmogorov-Smirnov (K-S) test. These subsets are shaded in red (NIR) and
blue (radio) filled colors in Figure~\ref{fig:histograms}. A K-S test comparing
the shaded portions of each dataset was performed. It revealed a 33\%
probability that the northern radio and NIR $\Delta {\rm P.A.}$ sets are drawn
from the same parent distribution. This K-S probability fell to 0.25\% for the
radio-NIR comparison in the south. Thus, in the north, the radio and NIR both
reveal a common \mbox{B-field} distribution (i.e., \emph{different}
distributions were not confidently detected). This is not the case in the south,
where the radio and NIR P.A. distributions are dissimilar at about the $3\sigma$
level.

\subsubsection{Radio-NIR Comparison along Major Axis}
\begin{figure}
{\centering
\includegraphics[angle=0, scale=1.2]
{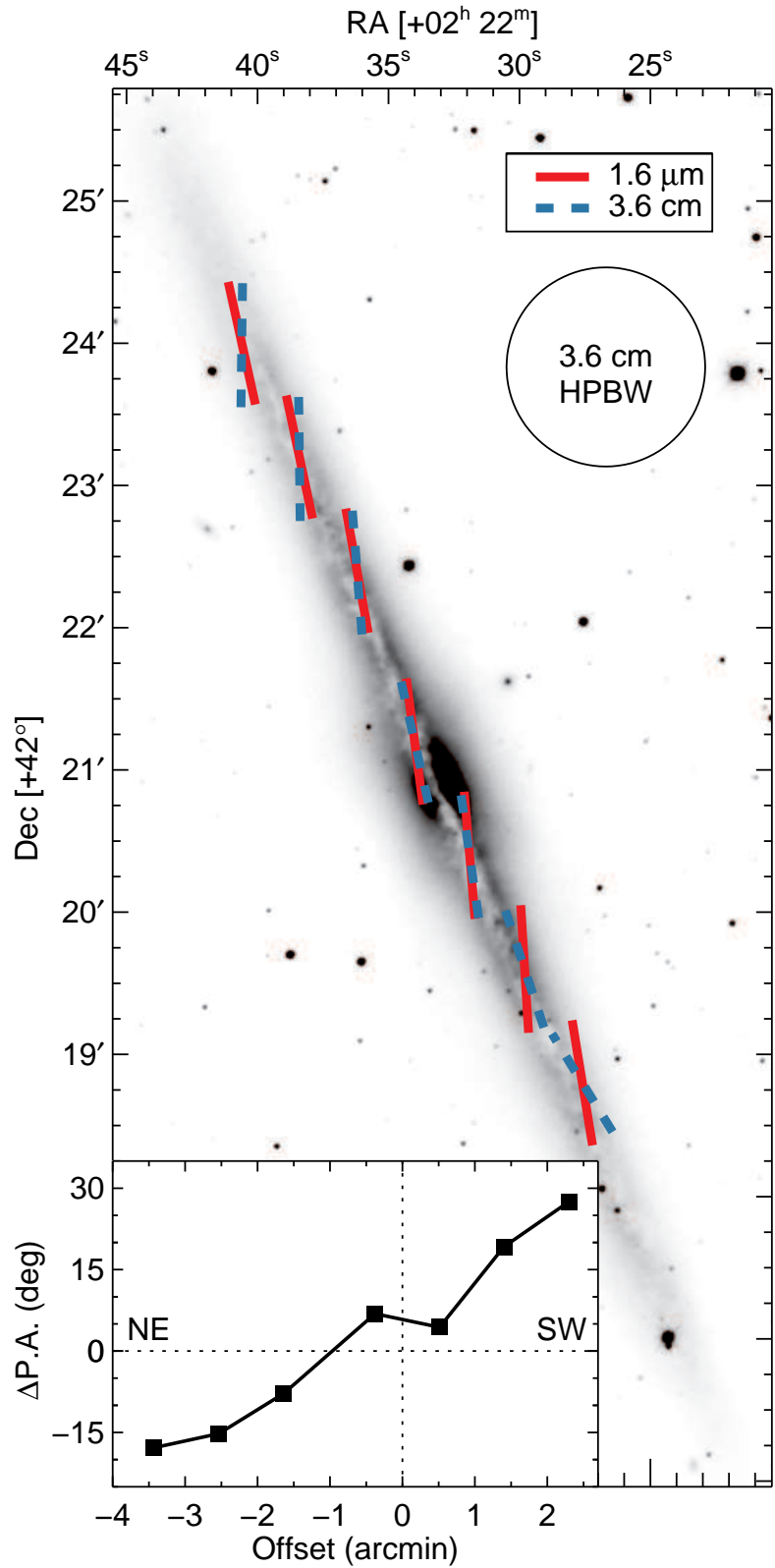}
\caption{
Vector representation of the radio and (beam-averaged) NIR P.A. values located
along the major axis. (inset:) Plot of $\Delta{\rm P.A.} = {\rm P.A.}_{radio} -
{\rm P.A.}_{NIR}$ as a function of angular offset from the galaxy center.
\label{fig:PAdiffPlot}}}
\end{figure}

Seven of the \citet{2009RMxAC..36...25K} radio P.A. measurements were centered
within a few arcseconds of the major axis. Gaussian (84\arcsec\ FWHM), inverse
variance weighted averages of $U$ and $Q$ were computed for all NIR data
obtained within the 84\arcsec\ HPBW radio beams centered on these points. Mean
$U$ and $Q$ values were converted to P.A. values using Equation~\ref{eq:pol}.
These radio and NIR P.A. values are plotted as colored vectors over the $H$-band
intensity image in Figure~\ref{fig:PAdiffPlot}. The inset of that figure shows
$\Delta{\rm P.A.} = {\rm P.A.}_{radio} - {\rm P.A.}_{NIR}$ as a function of
offset from the galaxy center along the major axis. The $\Delta{\rm P.A.}$
values near the galaxy center are smaller than the $\Delta{\rm P.A.}$ values in
the outer disk. Furthermore, there is a newly revealed trend of ${\rm
P.A.}_{radio} < {\rm P.A.}_{NIR}$ in the north, changing almost linearly to
${\rm P.A.}_{radio} > {\rm P.A.}_{NIR}$ in the south. In general, the radio
polarizations in the outer disk of NGC~891 turn upward to create the X-shape
morphology while the NIR polarizations do not.

\section{Discussion} \label{sec:discussion}
The new $H$-band polarization data presented in this study reveal generally
different P.A.s than seen for optical polarizations \citep{1996MNRAS.278..519S}.
In addition, the properties of the NIR null point(s) contradict the
\citet{1997ApJ...477L..25W} polarization model for an \mbox{edge-on} galaxy.
What causes these differences? Several observations indicate the presence of
extraplanar dust blown off the disk by supernovae and winds
\citep{2000AJ....119..644H, 2007ApJ...668..918B, 2009MNRAS.395...97W}. Do the
distributions of dust and starlight, or the presence of spiral arms, influence
the observed polarizations?

The new $H$-band polarization data revealed several north-south (N-S)
asymmetries. These include the discovery of polarization null point(s) in the NE
disk but not SW disk, strong polarization through the NE and central dust lane
but weak polarization in the SW dust lane, disk-parallel P.A.s in the NE but
slightly offset P.A.s in the south, and better agreement between NIR and radio
P.A. values in the northern disk than in the southern disk.

\subsection{Galaxy Structure}
\subsubsection{Dust and Stellar Distributions}
NGC~891 has a significant extraplanar dust component with an estimated vertical
scale height of about 2~kpc \citep{2007A.A...471L...1K, 2009MNRAS.395...97W},
while the stellar distribution scale height is 0.35~kpc
\citep{1998A&A...331..894X}. Much of this extraplanar dust has visual
extinctions of order unity \citep[$\tau_{H} \approx 0.15$]{2000AJ....119..644H}.
The minor axis polarization percentage profile (Figure~\ref{fig:minorPro}.a)
shows a strong increase in polarization strength beginning about 0.35~kpc off
the dusty midplane. The relative strength and vertical height at which this
polarization trend begins indicate a likely origin in disk starlight scattered
by the extraplanar dust and into the line-of-sight. Thus, toward the disk
midplane, where there is significant background starlight that can be polarized
by dichroism, NIR polarizations likely do trace the \mbox{B-field}. However, far
off the mid-plane, where there is little background starlight, polarizations are
likely dominated by scattering and do not trace the \mbox{B-field}. The regions
where scattering likely dominates the polarization are marked in
Figure~\ref{fig:minorPro} by gray shading.

\subsubsection{Spiral Arms}
The N-S asymmetries noted above may be related to spiral arms. Many other N-S
asymmetries were previously observed in NGC~891. \citet{1998A&A...331..894X} and
\citet{2007A.A...471L...1K} observed asymmetries in the NIR brightness and
H$\alpha$ light, respectively, and both concluded these indicated the presence
of a spiral arm on the near side (relative to the center of NGC~891) of the NE
disk. However, similar asymmetries in the \ion{H}{1} emission
\citep{1997ApJ...491..140S} and at other radio wavelengths
\citep{1994A&A...290..384D} suggest that spiral structure and extraplanar dust
might not be responsible for the observed asymmetries. If spiral arms influence
the observed distribution of light even in \mbox{edge-on} galaxies, where they
cannot be directly observed, then the polarization asymmetries---especially the
null points--- may be similarly related to the influence of spiral arms.
Section~\ref{sec:spiralNulls} further discusses the possible connection between
spiral arms and polarization null points.

\subsection{Dynamo and Polarization Models}
\subsubsection{$\alpha\Omega$ Dynamo}
The new NIR polarization data reported here confirm a significant difference
between the galaxy major axis P.A. and the $H$-band polarization P.A., as seen
toward the galaxy central region by \citet{1997AJ....114.1393J}, across most of
the full extent of NGC~891. This P.A. offset is at odds with the predictions of
typical $\alpha\Omega$~dynamo models \citep{1988Natur.336..341R}. Also, the P.A.
offset cannot be due to polarization effects within the Milky~Way, as the
$H$-band foreground extinction to NGC~891 is only 0.029~mag
\citep{2011ApJ...737..103S}. Interestingly, \citet{2012ApJS..200...21C} found
that the median Galactic Position Angle (GPA) for background starlight
polarizations in the Galactic plane is 75\arcdeg, a 15\arcdeg\ offset from the
expected disk-parallel GPA of 90\arcdeg. Thus, it is possible that significant
departures from toroidal B-fields are not uncommon.

\subsubsection{Polarization and Radiative Transfer}
\citet{1997ApJ...477L..25W} modeled the expected polarization due to scattering
and dichroism for a galaxy with purely toroidal \mbox{B-fields} for several
galaxy inclination angles. In his model, light from the bright galaxy center was
scattered by dust in the disk to produce disk-perpendicular polarizations
throughout the galaxy \citep[Figure~1]{1997ApJ...477L..25W} while light from the
stellar disk passed through magnetically aligned dust grains to produce
disk-parallel polarizations. In the emergent polarization images of the model
\mbox{edge-on} galaxies, a symmetric pair of polarization null points appeared
in the model disk at the projected offsets from the galaxy center where the
orthogonal scattering and dichroic polarizations canceled. The null points were
associated with a transition from disk-parallel polarization in the central
region to disk-perpendicular polarization further from the bulge.

The data presented in Figure~\ref{fig:majorPro} of the present study show at
least one highly significant NIR polarization null point, yet the polarization
orientation revealed in Figure~\ref{fig:regA} is disk-parallel on \emph{both}
sides of that null point. This result argues against any transition from
dichroic to scattering polarization across this null point.

The \citet{1997ApJ...477L..25W} model did not include spiral structure in the
distribution of starlight or dust or in the magnetic field geometry. This simple
model predicted the occurrence of a \emph{symmetric pair} of null points in the
disk, one on either side of the galaxy center. However,
Figure~\ref{fig:majorPro} shows that no null points are observed in the SW disk.
Rather, there is a continuous strip of low polarization along the SW disk dust
lane. Thus, the observed null point(s) are different in nature than those
predicted by the \citet{1997ApJ...477L..25W} model.

\citet{1997AJ....114.1405W} specifically modeled the galaxies NGC~4565 and
NGC~891 at $V$-band and $H$-band wavelengths, using the
\citet{1997ApJ...477L..25W} formalism, and they compared the model results to
existing polarization observations in these bands. They found that observed
polarizations of NGC~891 in both $V$- and $H$-band could not be replicated by
their model. They attributed the discrepancies to strong winds dragging the
\mbox{B-field} up into the halo, as evidenced by the detection of extraplanar
dust blown off the disk \citep{1997AJ....114.2463H}.

\subsubsection{Weak Polarization Percentage}
NGC~891 shows particularly weak $P$ across its entire disk. The JKD model
predicts $\sim 2\%$ polarization based on the $E(H-K)$ color of the dust lane of
NGC~891. \citet{1997AJ....114.1393J} noted that NGC~4565, another normal
\mbox{edge-on} galaxy, shows polarizations much stronger than NGC~891 and in
agreement with the JKD model. He suggested two possible explanations. The low
$P$ might be due to a crossed-polaroid effect from toroidal B-fields in the
inner disk and poloidal B-fields in the outer disk.
Section~\ref{sec:opticalVsNIR} discusses why this is unlikely. The other
\citet{1997AJ....114.1393J} explanation requires tangled B-fields on scales
smaller than 100~pc but not on any larger scales. This B-field configuration
seems untenable given Figures~4, 5, and 6 here.

A third explanation relates to scattered bulge light. Scattering of bulge light
by dust in the disk mid-plane is not the dominant source of polarization along
the major axis, although some scattered light ought to be present. This
scattered light would be polarized with disk-perpendicular P.A. and could
provide a crossed-polaroid effect, which would diminish the net polarization
percentage. However, the bulge to disk luminosity ratio of NGC~891 is not
particularly high, so it is unclear why this would be the case in this galaxy
but not in others. The low $P$ for NGC~891 is yet to be successfully explained.

\subsection{Polarization Null Points and Spiral Arms} \label{sec:spiralNulls}
If the polarization null points found here are not caused by the mechanism
described by \citet{1997ApJ...477L..25W}, then what \emph{does} cause null
points, and what influences their locations? Could the null points be related to
spiral arms? \citet{2011MNRAS.412.2396F} observed that the radio synchrotron
traced \mbox{B-field} is oriented parallel to the spiral arms of M51 (a
\mbox{face-on} galaxy). If the \mbox{B-field} in the \mbox{edge-on} NGC~891 is
similarly directed along its spiral arms, then for lines-of-sight with
significant optical depth inside of a spiral arm, the net plane-of-sky component
of the \mbox{B-field} will be small, and the resulting dichroic polarization
will be weak. Assuming a smooth exponential distribution of dust in the disk,
with a conservative $V$-band optical depth $\tau_{V} = 3$ through the galaxy
pole, model $H$-band optical depths were estimated. With a nearly \mbox{edge-on}
galaxy inclination, unit optical depth at $H$-band corresponds to a path length
of $\sim$6~kpc through the disk. Thus, depending on the scale lengths of the
dust distribution, $H$-band observations may probe deeply enough to reach points
where the ``magnetic spiral arms'' are tangent to the line-of-sight, producing
polarization nulls having disk-parallel polarization signatures flanking each
null.

\subsection{Optical vs. NIR polarization} \label{sec:opticalVsNIR}
\citet{1996MNRAS.278..519S} interpreted their observed $V$-band, predominantly
disk-perpendicular, polarizations as evidence for poloidal \mbox{B-fields} in
the outer disk of NGC~891. However, the scattering only, \mbox{edge-on} galaxy
in the \citet{1997ApJ...477L..25W} models produced disk-perpendicular
polarizations (even in the central region, where
\citeauthor{1996MNRAS.278..519S} argued scattering could not be responsible for
the observed polarization P.A.). Furthermore, if the
\citeauthor{1996MNRAS.278..519S} disk-perpendicular polarizations were caused by
poloidal \mbox{B-fields} in the outer disk, then the $H$-band polarizations
presented here for the extreme NE and SW disk should also have been
disk-perpendicular, but they are not. Instead, bulge light scattered by the
disk dust into the line-of-sight might be the dominant source of optical
wavelength polarized light, in which case optical polarization is unlikely to be
tracing the \mbox{B-field} of NGC~891.

\citet{2012ApJ...761L..28P} found no $H$-band polarization from the
\mbox{face-on} galaxy M51. If the lack of polarization was due to insufficient
NIR optical depth through the galaxy to produce detectable polarization, then
there may be similarly insufficient dichroic optical depth in the visible bands,
assuming a normal Serkowski law relationship \citep{1975ApJ...196..261S,
1982AJ.....87..695W} between polarization and wavelength. If so, the
centro-symmetric optical polarizations observed in \mbox{face-on} galaxies
\citep{1987MNRAS.224..299S, 1992MNRAS.257..309D, 1993MNRAS.264L...7S,
1996QJRAS..37..297S} may also be caused by scattering and not dichroism. This
would leave radio as the sole means for probing the \mbox{B-field} of
\mbox{face-on} galaxies and weaken the reliability of optical polarizations for
probing the \mbox{B-fields} of even \mbox{edge-on} galaxies.

\section{Summary}
Deep $H$-band imaging polarization observations of the Milky~Way analog
\mbox{edge-on} galaxy NGC~891 revealed moderately weak ($P < 1\%$) mean
polarization, though detected with SNR up to 20. These data were sensitive
enough to permit detection of NIR polarization in the disk of NGC~891 out to
14~kpc from its center, which is much further than any previous NIR polarimetric
observations. The polarization map revealed at least one polarization null point
in the NE disk but none in the SW disk. The northern inner null point does not
show the expected transition from disk-parallel to disk-perpendicular
polarization with increasing projected central offset
\citep{1997ApJ...477L..25W}. NIR polarization P.A. for NGC~891 is generally
oriented 10\arcdeg\ to 20\arcdeg\ west of the galaxy major axis P.A., even far
from the central bulge. The wide angular extent of these data permitted
comparison between NIR and radio synchrotron polarizations. In general, the
distribution of radio and NIR P.A. values are more similar in the north than in
the south, indicating that radio and NIR are generally probing a common
\mbox{B-field} in the north but different \mbox{B-fields} in the south. A trend
in $\Delta{\rm P.A.} = P.A._{radio} - P.A._{NIR}$ along the major axis and
several other N-S asymmetries in the polarization may be related to spiral
structure in the disk.

Polarization in the dust lane showed several distinct features. Polarization
percentage reaches a maximum in the dust lane for the NE disk and central
regions but exhibits a minimum in the SW disk dust lane. The P.A. values shift
from disk-parallel east of the dust lane to $-20$\arcdeg\ (relative to the major
axis) west of the dust lane. Polarization percentage increases with distance
from the disk mid-plane. This result and the disk-parallel P.A.s above and below
the NE disk, may be caused by disk starlight scattered by extraplanar dust.

NIR polarizations exhibit distinctly different orientations than seen at optical
wavelengths. There is no disk-perpendicular NIR polarization in the outer disk,
as would be expected if poloidal \mbox{B-fields} were present. Rather, there is
very weak NIR polarization in the far NE disk, and the polarizations in the far
SW disk are even more disk-parallel than elsewhere in the galaxy. It seems
likely that the optical polarizations in this galaxy are dominated by scattering
and that these new NIR polarization observations (which generally agree with
radio synchrotron values) better reveal the magnetic field orientations in the
cool ISM of NGC~891.

\acknowledgments

Private communication with A. Schechtman-Rook regarding spiral structure in
galaxy disks was greatly appreciated. This research used SAOImage DS9, developed
by the Smithsonian Astrophysical Observatory. This publication makes use of data
products from the Two Micron All Sky Survey, which is a joint project of the
University of Massachusetts and the Infrared Processing and Analysis
Center/California Institute of Technology, funded by NASA and the NSF. This work
is based in part on observations made with the \textit{Spitzer Space Telescope}
and hosted by the NASA/ IPAC Infrared Science Archive which are both operated by
the Jet Propulsion Laboratory and California Institute of Technology under a
contract with the NASA. This research was conducted in part using the Mimir
instrument, jointly developed at Boston University and Lowell Observatory and
supported by NASA, NSF, and the W.M. Keck Foundation. Analysis software for
Mimir data was developed under NSF grants AST 06-07500 and 09-07790 to Boston
University. Grant AST~09-07790 from NSF to Boston University and grants of
observing time from Boston University-Lowell Observatory partnership are
gratefully acknowledged.

{\it Facility:} \facility{Perkins (Mimir)}

\clearpage

\clearpage

\clearpage

\clearpage

\clearpage

\clearpage

\clearpage

\clearpage

\end{document}